\documentclass[aip,jcp,reprint]{revtex4-2}
\usepackage{graphicx}
\usepackage{xcolor}
\usepackage{amsmath}
\newcommand{\sub}[1]{\ensuremath{_{\rm #1}}} 
\newcommand{\super}[1]{\ensuremath{^{\rm #1}}} 
\renewcommand{\vec}[1]{\ensuremath{\boldsymbol{\mathbf{#1}}}}
\newcommand{\edited}{\textcolor{black}}
\newcommand{\affilRPIMSE}{\affiliation{Department of Materials Science and Engineering, Rensselaer Polytechnic Institute, Troy, NY 12180, USA}}
\newcommand{\affilRPIPhy}{\affiliation{Department of Physics, Applied Physics, and Astronomy, Rensselaer Polytechnic Institute, Troy, NY 12180, USA}}
\newcommand{\affilSC}{\affiliation{Department of Chemistry and Biochemistry, University of South Carolina, Columbia, SC 29208, USA}}

\begin{document}
\title{Bridging electronic and classical density-functional theory\\
using universal machine-learned functional approximations}
\author{Michelle M. Kelley}\email{kellem6@rpi.edu}\affilRPIMSE
\author{Joshua Quinton}\affilRPIPhy
\author{Kamron Fazel}\affilRPIMSE
\author{Nima Karimitari}\affilSC
\author{Christopher Sutton}\affilSC
\author{Ravishankar Sundararaman}\email{sundar@rpi.edu}\affilRPIMSE\affilRPIPhy

\begin{abstract}
The accuracy of density-functional theory (DFT) calculations is ultimately determined by the quality of the underlying approximate functionals, namely the exchange-correlation functional in electronic DFT and the excess functional in the classical DFT formalism of fluids.
For both electrons and fluids, the exact functional is highly nonlocal, yet most calculations employ approximate functionals that are semi-local or nonlocal in a limited weighted-density form.
Machine-learned (ML) nonlocal density-functional approximations show promise in advancing applications of both electronic and classical DFTs, but so far these two distinct research areas have implemented disparate approaches with limited generality.
Here, we formulate a universal ML framework and training protocol to learn nonlocal functionals that combines features of equivariant convolutional neural networks and the weighted-density approximation.
We prototype this new approach for several 1D and quasi-1D problems and demonstrate that functionals with \emph{exactly the same hyperparameters} achieve excellent accuracy for a diverse set of systems including the hard-rod fluid, the inhomogeneous Ising model, the exact exchange energy of electrons, the electron kinetic energy for orbital-free DFT, as well as for liquid water with 1D inhomogeneities.
These results lay the foundation for a universal ML approach to approximate exact 3D functionals spanning electronic and classical DFTs.
\end{abstract}
\maketitle

\section{Introduction}

Electronic density-functional theory (DFT) has become indispensable in physics, chemistry, and materials science, with its low cost and respectable accuracy for electronic structure predictions making it a vital component of tens of thousands of scientific publications each year.\cite{BurkeDFTperspective}
The accuracy of electronic DFT hinges on the quality of the approximate exchange-correlation functionals, and advances from local through semi-local to nonlocal functionals have gradually improved the reliability and applicability of DFT.\cite{DFTLadder}
In the meantime, machine learning (ML) has shown promise for accelerating these improvements further, to more rapidly approach the in-principle exact DFT functional.\cite{DFT-ML-Burke, BypassKS}

As ubiquitous as electronic DFT has become as an electronic structure method, DFT itself is actually a much more general technique widely applicable to many-body interacting systems beyond just systems of electrons.
In particular, classical DFT is a statistical mechanics technique capable of describing equilibrium properties of inhomogeneous fluids in terms of their density distributions alone, eliminating the need to thermodynamically sample the intractably large configuration space of fluid molecules.\cite{Mermin-DFT, Capitani-UDFT}
This approach is particularly appealing for solvation, specifically combining classical DFT with electronic DFT to capture compounding effects between solvents and electrolytes with the electronic structure of the solutes.\cite{JDFT}

Developing functional approximations for classical DFT, however, is fundamentally more challenging than for the electronic case for two reasons.
First, there is precisely one exact exchange-correlation functional that needs approximation in electronic DFT, which can then be used for calculations of any material or any conceivable configuration of atoms.
In contrast, classical DFT requires an excess functional approximation to be constructed, painstakingly, for each individual liquid or electrolyte of interest.
Second, semi-local formulations of approximate functionals, which are reliably successful in electronic DFT, are grossly inadequate for approximating excess functionals of liquids.
Minimum viable classical DFTs are nonlocal, starting at the weighted density approximation (WDA) level,\cite{WDA} extending to the rank-2 decomposition and fundamental measure theory approach,\cite{PercusRank2, FMTreview} and eventually perturbed to develop approximations for real fluids.\cite{WuIntegralEqnCDFT, WuBonding, BondedTrimer, RigidCDFT, PolarizableCDFT}
Accordingly, the complexity of developing the necessary nonlocal functionals for liquids has limited the widespread application of classical DFT. 
On the other hand, ML offers a solution to overcoming this current bottleneck to facilitate the development of such functionals,\cite{Wu2023} with promising results for model fluids \edited{such as hard disks and hard spheres}.\cite{InhomogeneousFluidNN, NeuralFunctionalsStatMech}

Machine-learning has had a significant impact in the materials and electronic structure fields, ranging from rapid property predictions bypassing DFT entirely, to bypassing solving the Kohn-Sham equations either through the direct prediction of DFT quantities (e.g., electron density),\cite{BypassKS, Grisafi2019} or through predicting the Hamiltonian matrix.\cite{Nigam2022, unke2021seequivariant, Zhong2024, liDeeplearningDensityFunctional2022, gongGeneralFrameworkEquivariant2023}
Moreover, several ML approaches have been implemented to improve the DFT functionals by developing better semi-local and nonlocal exchange-correlation functionals,\cite{ML-metaGGA-HeadGordon, NeuralXC, bystromNonlocalMachineLearnedExchange2023} as well as for the kinetic energy functional for orbital-free DFT.\cite{DFT-ML-Burke, delmazo-sevillanoVariationalPrincipleRegularize2023}

The fundamental differences between the underlying interactions encompassed in the traditional electronic and classical DFTs require the two methods to adopt distinct approaches for functional development.
Correspondingly, the applications of ML for functional development have also been disjoint in these two fields, despite having the same fundamental goal: approximating a functional that maps a density distribution to the energy.
Therefore, a unified approach to accelerate functional developments across both fields is highly desirable.

Here, we introduce a single universal approach for developing nonlocal functionals across a wide range of fields, spanning both electronic and classical DFT.
Below, we will first briefly describe density functionals in both quantum and classical contexts.
Then, we emphasize on our \edited{universal} framework combining equivariant convolutional neural networks with the weighted-density approximation and establish a general training protocol across different classes of DFTs.
Finally, we will apply this universal technique \edited{and build functionals for} the 1D hard-rod fluid, the Ising model, Hartree-Fock exchange, the Kohn-Sham kinetic energy and liquid water, and demonstrate excellent accuracy across all of these cases with the same hyperparameters.

\edited{We intentionally select this diverse set of classical and quantum systems to include both simple and complicated examples to demonstrate the universality of our approach.
Specifically, for classical DFT, the hard-rod fluid with an exact functional is the simple limit, while the highly-complex response of liquid water tests the capabilities of the approach.
Similarly, Hartree-Fock exchange is the easier test for the electronic case, because the orbital-dependent kinetic energy helps regulate the total energy, whereas the Kohn-Sham kinetic energy lacks this regulation and serves as the more difficult functional to learn.}
The software used to produce and test all of the machine-learned density functionals below is available in an open-source git repository (see the Supplementary Information for more details).

\section{Theoretical Formulation}

Density-functional theory is a general theorem that establishes the ground-state energy $E$, or similarly the equilibrium free energies, of a many-body system in an external potential $V(\vec{r})$ can be obtained by minimizing a functional of the density $n(\vec{r})$ alone, as
\begin{equation}
E = \min_{n(\vec{r})} \left[ F[n(\vec{r})] + \int d\vec{r} V(\vec{r}) n(\vec{r}) \right],
\end{equation}
where $F[n]$ is a universal functional of the density, independent of the potential $V$.
Applied to electrons, $F$ is the Hohenberg-Kohn energy functional in the microcanonical ensemble,\cite{HK-DFT} or the Mermin Helmholtz free-energy functional in the canonical ensemble.\cite{Mermin-DFT}

This theorem only requires that the interaction of the many-body system with the external potential takes the form $\int d\vec{r} \,V(\vec{r}) n(\vec{r})$, and does not depend on the details of the internal interactions of the system.
Applied to a classical fluid with atomic densities $n_\alpha(\vec{r})$ in external potentials $V_\alpha(\vec{r})$ and chemical potentials $\mu_\alpha$ (in the grand-canonical ensemble), with $\alpha$ indexing species, \textit{e.g.}, O, H for water, the equilibrium grand free energy is
\begin{equation}
\Phi = \min_{n_\alpha(\vec{r})} \left[ F[n_\alpha(\vec{r})] + \sum_\alpha\int d\vec{r} (V(\vec{r}) - \mu_\alpha) n_\alpha(\vec{r}) \right].
\end{equation}

Practical applications of DFT require an approximation of the unknown exact functional $F$.
In both electronic and classical DFT, the standard approach is to break out known exact pieces, leaving behind the smallest and hopefully easiest piece to approximate.
In electronic DFT in the Kohn-Sham formalism,\cite{KS-DFT} the exact functional is split as 
\begin{equation}
F[n] = T_S[\{\psi_i[n]\}] + E_\textrm H[n] + E_\textrm{xc}[n],    
\end{equation}
separating out the exact orbital-dependent kinetic energy of the non-interacting electronic system $T_S$ at the same density $n(\vec{r})$ and the mean-field Coulomb interaction $E_\textrm H$, leaving behind the exchange-correlation functional to approximate.
Furthermore, direct approximation of $T_S[n]$ from the electron density would enable orbital-free DFT, however it is generally very difficult to achieve accuracy comparable to the orbital-dependent version in Kohn-Sham theory.
Correspondingly, classical DFT typically partitions the exact functional as
\begin{equation}
F[n_\alpha(\vec{r})] = F\sub{id}[n_\alpha(\vec{r})] + F\sub{ex}[n_\alpha(\vec{r})],    
\end{equation}
separating out the exact ideal gas free energy $F\sub{id}$ at the same set of densities, leaving behind the excess functional $F\sub{ex}$ that needs to be approximated.

Conventional approaches to approximate $E_\textrm{xc}[n]$, $T_S[n]$ and $F\sub{ex}[n_\alpha]$ include semi-local approximations of the form $F[n] = \int d\vec{r}\, n f(n, \nabla n)$ and the weighted density approximation (WDA) $F[n] = \int d\vec{r}\, n f(w \ast n)$, where the convolution by a weight function $w$ introduces nonlocality.
Semi-local approximations are widely applied for $E_\textrm{xc}$, but much less successful for $T_S[n]$ and grossly inadequate for $F\sub{ex}$, where at least the WDA is necessary.

\subsection{Machine-learning density functionals}
Here, we propose universal ML density-functional approximations as a sequence of convolution ($C$) and activation ($A$) layers that ultimately culminates in a readout ($R$) layer,
\begin{equation}
\begin{split}
F[n_\alpha(\vec{r})]
= \int d\vec{r} \,R(n_\alpha,\, &C_{\alpha_m}[
A_{\alpha_{m-1}'}(C_{\alpha_{m-1}}[\\
&\cdots
A_{\alpha_{1}'}(C_{\alpha_{1}}[
n_\alpha(\vec{r})
])])]).
\end{split}
\end{equation}
Here, $\alpha$ label channels in the input and through all the intermediate layers.
At the input, these channels are physical indices of the density, such as spin for electrons or atomic site type for a classical fluid (\textit{e.g.} $O, H$ for water).
The intermediate $\alpha_i$ label the output channels of each $C$ or $A$ layer, with the number of these layers and the number of $\alpha_i$ at each intermediate layer being hyperparameters of the model.
Note that the readout $R()$ and activation $A()$ are functions that operate locally in space, while $C[\ ]$ is a functional that is nonlocal in space.

Each input or intermediate $\alpha$ index corresponds to a specific representation of the rotation group of space. In 1D, these indices corresponds to odd or even representations of the $Z_2$ group (reflection of the single axis).
While in 3D, these indices would correspond to $l$ and $m$ of the spherical harmonics\edited{, and in 2D, they would correspond to $m$ of the circular harmonics.}
To rapidly prototype and test the universality of the ML DFT approach across several distinct problems, we focus \edited{the software implementation of this approach and its applications} here on 1D systems.
The overall structure and logic extends straightforwardly to the \edited{2D and 3D cases}, as we will point out briefly for each operator below.

\subsubsection{Convolution layers}
We define the linear convolution layers as
\begin{equation}
C_\alpha[n_\beta](\vec{r}) =
\int d\vec{r}' w_{\alpha\beta}(\vec{r} - \vec{r}')
n_\beta(\vec{r}')
\end{equation}
with learnable weight functions $w_{\alpha\beta}$.
Each input density index $\beta$ corresponds to an even or odd function, and, likewise, each output channel index $\alpha$ is also either even or odd.
Correspondingly, each weight function $w_{\alpha\beta}$ is even if $\alpha$ and $\beta$ belong to the same parity and is odd otherwise.
For the 3D case, the mapping from inputs to outputs involves Clebsh-Gordon coefficients to combine the $(l,m)$ and $(l',m')$ of the inputs and weights into $(L, M)$ of the outputs.
\edited{For the 2D case, the angular index adds as $M=m+m'$.
Unlike the finite rotation group for the 1D case, the rotation groups in both 2D and 3D are not finite, and additionally require truncation at finite $m$ and $l$ respectively in practice.}
The rest of the structure remains unchanged.

To preserve rotational invariance of the energy, the final convolution layer must output only even-weighted densities (or $L=0$ in 3D) for the WDA-style readout function discussed below.
See the Supplementary Information (SI) for an alternate readout function motivated from the `rank-2 approximation' form proposed by Percus for classical DFT,\cite{PercusRank2} which can handle non-scalar weighted densities as inputs.
Our numerical tests indicate that the WDA-style readout works more generally, so we focus on this approach here.

The learnable parameters of the convolution layer are within the weight functions, which we define as smooth functions $\tilde{w}(G)$ in reciprocal space.
This is in contrast to typical convolution neural networks, where the weights are discrete and correspond to a certain number of nearest neighbors.
Instead, the $\tilde{w}(G)$ definition allows for readily porting the functional between variable grid spacings, dimensions, and geometries which is imperative for both training and applying the model.

We implemented and tested several parameterizations  of the weight functions, including cubic splines, neural networks and Gaussians multiplied by polynomials (see the SI for details).
We find the best performance using Gaussians multiplied by a polynomial of degree $d$,
\begin{equation}\tilde{w}(G) =
\exp\left(-\frac{(\sigma G)^2}{2}\right)
\sum_{n=0}^{d} a_n (\sigma G)^{2n},
\label{eq:gaussweights}
\end{equation}
where $\sigma$ and $\{a_i\}$ are learnable parameters.
In particular, as we show below, degree $d=1$ appears to provide the best balance between flexibility of the weight function shape and the trainability of the overall functional.
We implement learnable even weight functions in reciprocal space and build the corresponding odd weight functions by introducing an extra factor of $iG$.

\subsubsection{Activation layers}
The activation layers between adjacent convolution layers introduce nonlinearity, which is necessary because otherwise, two adjacent convolution layers would trivially reduce to one convolution layer.
We define the activation layer as
\begin{equation}
A_\alpha(n_\beta) =
n_\alpha \Theta\left[ b_\alpha +
\sum_{\beta|n_\beta\ even} W_{\alpha\beta} n_\beta
\right],
\end{equation}
where the learnable parameters are the biases $b_\alpha$ and the weights $W_{\alpha\beta}$, are denoted with a capital to distinguish from the weight functions in the convolution layer.
Importantly, the nonlinear functions must use only the invariants (scalars) as inputs in order to preserve rotational symmetry.
Hence only the $\beta$ corresponding to even weighted-densities (or $l=0$ in 3D) are used as inputs.
The activation function $\Theta$ can be any smooth activation function used for neural networks, and we use the SoftPlus form here, $\Theta(x) = \ln(1 + e^x)$.
Essentially, each input passes through to the output multiplied by an activation function $\Theta$ applied to a learnable linear combination of \emph{only the scalar} inputs to preserve equivariance.

\subsubsection{Readout layer}
The readout function predicts the energy density from the original input site densities $n_\alpha(\vec{r})$ and the outputs of the final convolution layer $\bar{n}_{\alpha_M}(\vec{r}) = C_{\alpha_M}[\cdots C_{\alpha_1}[n_\alpha]]$.
We use the weighted density notation, $\bar{n}$, for these outputs as the densities are weighted in the style of the weighted-density approximation for the case of one single convolution layer.
When there are multiple convolution layers with intervening activation layers, these layers collectively implement a nonlinear convolution.
This nonlinear convolution adjusts the range and shape of the weight functions based on the local environment.

Writing $\bar{n}_\beta\equiv\bar{n}_{\alpha_M}$ for brevity, the readout function fits the energy-per-particle as
\begin{equation}
 R(n_\alpha, \bar{n}_\beta) = 
\sum_\alpha n_\alpha f_{\alpha}(\bar{n}_\beta, T, \ldots),
\label{eq:readout}
\end{equation}
where $f_{\alpha}$ is a multi-layer perceptron (MLP) that takes all the scalar-weighted (even, $l=0$) densities $\bar{n}_\beta$ and any global attributes, such as the temperature $T$, together as a single input vector.
(See the SI for an alternate `rank-2' formulation\cite{PercusRank2} that also admits non-scalar inputs.)
For the Kohn-Sham case, we find that additionally providing the input densities $n_\alpha$ as a direct input to $f$ leads to better results. 
Otherwise, the network will attempt to learn local density information by making one or more weight functions sharp which can lead to numerical issues in the subsequent Euler-Lagrange minimization of the functional.
The hidden layer sizes and choice of activation function for the MLP are additional hyperparameters to the model.

\subsubsection{Loss function and training protocol}
\label{sec:loss+training}

We train the ML functionals to a set of corresponding densities $n_i(r)$, value of the unknown functional $F_i$ and functional derivative $\delta F_i / \delta n$, evaluated from the exact or reference system in an assortment of external potentials, as described below in Section~\ref{sec:datageneration}.
We then minimize the loss function
\begin{equation}
\begin{split}
L=\sum_i& \Biggr[c_E \left(F[n_i]-F_i\right)^2 \\
&+ c_V \int \frac{d\vec{r}}{\Omega_i} \left( \left.\frac{\delta F}{\delta n(\vec{r})}\right|_{n_i(\vec{r})} - \frac{\delta F_i}{\delta n} \right)^2 \Biggr],
\end{split}
\label{eq:loss}
\end{equation}
where $\Omega_i$ is the volume of the domain for reference calculation or simulation $i$.
Above, $c_E$ and $c_V$ allow balancing the relative influence of the energy and potential on training the functional.
In principle, these could be adjusted dynamically during training to accelerate optimization of the parameters, but we find that $c_E = c_V = 1$ leads to adequate training performance for all of our test cases.

Our results below use a common set of hyperparameters, denoted as the \edited{`universal model' in each case; note that this signifies a separate model trained to distinct data for each physical problem, but with \emph{universal hyperparamters}}.
For the readout, the universal model employs three layers with 100 neurons in each layer.
There are three convolution layers in the model with 10 odd and 10 even weight functions in the first two layers and 20 even weight functions in the last convolution layer to preserve rotational invariance.
We use Gaussian weight functions with learnable $\sigma \le 4$ (length units, bohrs for the electronic case) and multiplied by polynomials of degree one.
In addition, for each application, \edited{we test a model with specialized, often reduced, set of hyperparameters} specific to that application, and compare its performance with the universal-hyperparameters model mentioned above.
These hyperparameters are reported in Table~\ref{tab:hyperparams} and discussed at the end of the Results section.

\section{Data generation}
\label{sec:datageneration}

\subsection{Hard-rod fluid}
For our first test case, we begin with the 1D hard-rod fluid for which an analytic density functional exists,\cite{Percus1976} \edited{where we expect the ML functional to certainly perform well}.
The excess functional for hard-rod fluid of length $a$ at temperature $T$ is exactly
\begin{equation}
F_\textrm{ex}[n] = -T\int dz\, n(z) \log \left(1-\int_{z-a}^z dz' n(z') \right),
\label{eq:hardrods}
\end{equation}
which is a nonlocal WDA form that happens to be exact for this system.
Note that we use Hartree atomic units throughout with $e, m_e, \hbar, k_B$ all set to 1 unless otherwise specified.
This solution can be generalized for hard-rod fluid mixtures\cite{Vanderlick1989} and for hard rods with contact nearest neighbor interactions.\cite{Percus1982}

We generate training data for the hard-rod fluid by solving the Euler-Lagrange equation using the conjugate-gradients algorithm in several random external potentials.
Without loss of generality we set the hard-rod length $a = 1$ and the temperature $T = 1$, as these parameters just control the overall length and energy scale respectively.
We generate random periodic potentials in reciprocal space as
\begin{equation}
\tilde{V}(G) =  \frac{\sigma\sqrt{2\pi}}{\Omega} r_G e^{-(\sigma G)^2/2},
\end{equation}
where $r_G$ are unit normal random numbers, $\sigma$ sets the length scale of variations (smoothness) of the potential and $\Omega$ is the length of the 1D unit cell.
The normalization factor sets the expectation value of $\int dx |V(x)|^2$ to 1.
We randomly select several bulk densities of the fluid (controlled by chemical potential $\mu$), $\Omega$ and $\sigma$, and for each case apply a sequence of potentials with increasing amplitude of the above random shape. 
(See the SI for a detailed description of the range of each of these parameters in the training data.)

\begin{figure}
\centering
\includegraphics[width=\columnwidth]{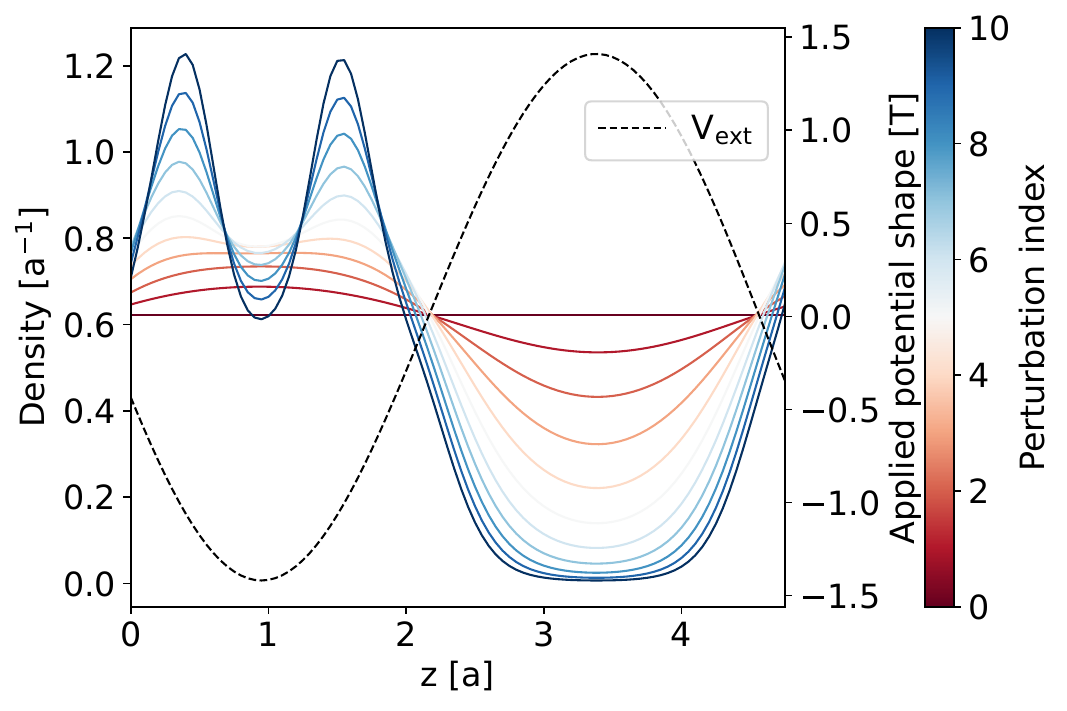}
\caption{Training data, consisting of the equilibrium densities in a sequence of perturbating potentials with the same shape (dashed line) and increasing amplitude (color scale), shown here for the hard-rod fluid.
The densities change from the bulk value at zero potential to a strongly inhomogeneous profile with density approaching zero in the repulsive regions of the potential and saturation with a shell structure in the attractive regions of the potential.
}
\label{fig:data_hardrod}
\end{figure}

Figure~\ref{fig:data_hardrod} shows a sample of the hard-rod fluid training data for one external potential shape.
The colored lines show the solutions of the local density profiles over a range of strengths of the external potential, gradually varying from the uniform fluid to a strongly inhomogeneous fluid with different regions approaching zero density and saturation.
The minimization of Eq.~\eqref{eq:hardrods} provides the equilibrium energies $E_i$ and densities $n_i(r)$, from which we can calculate $F_i$ and $\delta F_i/\delta n$ for the excess functional.
We generate 1000 such sets, split these data 80\%-20\% for training and testing, train an ML functional and check how well it reproduces the results from the known exact functional.
(See the SI for results on the inhomogeneous Ising model in 1D, which is an additional exactly-solvable case to examine the performance of the universal ML model.)

\subsection{Hartree-Fock exchange energy}

\edited{Substantially increasing the complexity of the physical problem, we next consider the exchange energy functional in electronic DFT.}
Unlike the previous example, there is no analytic density-functional for the exact exchange energy of the Hartree-Fock system.
Instead, the exact exchange energy is a nonlocal functional written in terms of the electronic orbitals $\psi_i(\vec{r})$,
\begin{equation}
E\sub{xx}=\frac{-1}{2}\sum_{ij}\iint d\vec{r} d\vec{r}'
\psi^*_i(\vec{r})\psi_j(\vec{r})\psi_i(\vec{r}')\psi^*_j(\vec{r}')
v\sub{c}(\vec{r}-\vec{r}'),
\label{eq:exx}
\end{equation}
where $v\sub{c}$ is the Coulomb kernel.
Note that in 1D, the Coulomb kernel must be regularized at short length scales to avoid a divergence and we use $v_c(z) = 1/\sqrt{z^2 + a^2}$ as proposed in previous studies of electronic exchange and correlation in 1D.\cite{SoftCoulomb1D}

The total energy of this Hartree-Fock system is calculated in a DFT framework using
\begin{equation}
E=T_S[\{\psi_i\}]+E\sub{xx}[\{\psi_i\}]+E_H[n] +\int d\vec{r} V(\vec{r}) n(\vec{r}),
\label{eq:dft-exx}
\end{equation}
where $T_S$ and $E_\textrm H$ are the kinetic energy and mean-field Coulomb (Hartree) terms as discussed above.
While the external and Hartree energies are explicit functionals of the electron density $n(\vec{r})$, the kinetic and exchange energies are orbital dependent.

Despite the orbital dependence of the above energy functional, one can still identify an effective local potential for a Kohn-Sham system 
\begin{equation}
\left(-\frac{\nabla^2}{2} + V_\textrm{eff}(\vec{r}) \right)\psi_i(\vec{r})=\epsilon_i \psi_i(\vec{r})
\label{eq:ks}
\end{equation}
that yields the exact solutions for the ground state energy and density $n(\vec{r})=\sum_i |\psi_i(\vec{r})|^2$ by applying the optimized effective potential (OEP) method.\cite{OEP_Wu2003}
Instead of the usual Kohn-Sham variational equations formulated for explicit density-functionals, the more general OEP method provides an analogous approach for orbital-dependent functionals.
We solve for the effective local potential $V_\textrm{eff}(\vec{r})$ by invoking the variational principle and minimizing the total energy with respect to variations in the potential such that $\delta E/ \delta V_\textrm{eff}(\vec{r})=0$, with the functional derivative of $E[\psi[V]]$ evaluated exactly using automatic differentiation in PyTorch, instead of the conventional perturbation theory approach.\cite{OEP_Krieger1992}

We generate training data for the Hartree-Fock exchange functional in smooth random periodic external potentials, as was done in the case of the hard-rod fluid.
In addition, we supplement the training data with random molecules, generated by sampling different numbers of atomic nuclei of various charges to be placed at various separations.
Note that the nuclear potentials also use the regularized Coulomb kernel $v_c(z) = 1/\sqrt{z^2 + a^2}$, consistent with the Hartree and exchange terms, and we set $a = 1$ bohr for all our tests.
For each case, we use the equilibrium density $n(\vec{r})$, the exchange energy $E\sub{xx}$ and its functional derivative, obtained from the OEP potential as
\begin{equation}
\frac{\delta E\sub{xx}}{\delta n(\vec{r})} = V\sub{eff}(\vec{r}) - V(\vec{r}) - \frac{\delta E_\textrm H}{\delta n(\vec{r})},
\end{equation}
for both the training and testing datasets.

\subsection{Kohn-Sham kinetic energy}

\edited{For an even more challenging test of the ML functional approach, we target the} Kohn-Sham kinetic energy $T_S[n]$, which has been attempted previously with specialized approaches.\cite{DFT-ML-Burke, BypassKS}
\edited{The complexity of this case stems from completely bypassing Schrodinger equation solutions and single-particle orbitals, with the ML functional tasked with predicting the total energy based on the electron density alone.}
Since the exact kinetic energy of the non-interacting Kohn-Sham system is orbital-dependent,
\begin{equation}
T[\{\psi_i\}]=-\frac{1}{2}\sum_i \langle \psi_i| \nabla^2 | \psi_i\rangle.
\label{eq:kinetic}
\end{equation}
we use the OEP approach to construct the density-based training data.
Specifically, we solve Eq.~\eqref{eq:ks} in various external potentials $V(\vec{r}) = V\sub{eff}(\vec{r})$ by directly diagonalizing the Hamiltonian in the plane-wave basis for several $k$-points and compute the corresponding electron density $n(\vec{r})$.
We then calculate the kinetic energy $T_s$ and its functional derivative $\delta T_s/\delta n = \mu - V$ from the OEP equation.
As above, we produce training data for several smooth random period potentials, spanning a range of box sizes, smoothness and perturbation amplitudes. 

\subsection{Liquid water}
\label{sec:waterdata}

Finally, we target a excess free energy functional for liquid water, \edited{which tests the capable of the ML approach to capture the complex (1D) response of a real (3D) liquid in classical DFT.}
A key distinction in this case is that the reference data is not generated from an exact energy functional, but from 3D molecular dynamics (MD) simulations.
Here, we restrict to 1D inhomogeneity and consider the planar-averaged density in order to test the universal functional on the same footing as the 1D problems considered above.

We use the mdext extension\cite{mdext} to LAMMPS\cite{LAMMPS} to perform MD simulations of water with the SPC/E interatomic potential\cite{SPCE} in the presence of external potentials.
We use a $30\times 30 \times 40$~\AA~ box with $\sim 1200$ water molecules in the NVT ensemble at $T=300$~K, equilibrated for 50~ps and collected for 100~ps at each potential strength.
In this case, we do not use the random periodic potential because of a challenge in the MD simulations: a potential with multiple minima could lead to trapping of molecules as the strength of the potential increases, with exponential slowdown of the exchange of molecules between the multiple wells formed.
To circumvent this MD equilibration issue, we use a potential with shape
\begin{equation}
V(z) = e^{-z^{2}/2\sigma^{2}} (1 + Bz^{2})
\begin{cases}
1, & B < 0\\
\pm 1, & 0 \le B \le 0.5\\
-1, & B > 0.5
\end{cases},
\end{equation}
with randomly selected $B$ and $\sigma$.
The sign ensures that the potential produces at most one minimum (for $z > 0$) and avoids the multiple well problem. 
(This parameterization could lead to two symmetric wells in the MD simulation, but they will end up with nearly the same number of molecules when the trapping occurs, which is the correct equilibrium condition.)

We perform a sequence of simulations with increasing strength of the potential $\lambda V(z)$ applied to the O atoms and collect the density $n_\lambda(z)$ of the O atoms of water.
This sequence is a necessity to calculate the free energy of water by thermodynamic integration,
\begin{equation}
A_\lambda = \Omega a_0 + \int_{0}^{\lambda} d\lambda' \int dz V(z) n_{\lambda'}(z),
\end{equation}
where $a_0$ is the bulk free energy density of the liquid.
The corresponding functional derivative $\delta A_\lambda/\delta n(z) = -\lambda V(z)$ by the Euler Lagrange equation.
Finally, we subtract the total ideal gas free energy in the external potential, $\int dz\, n (T(\log n - 1) + \lambda V - \mu)$, and its corresponding functional derivative contribution, to generate the $F$ and $\delta F/\delta n$ required to train the excess functional.

\begin{figure}[t]
\includegraphics[width=\columnwidth]{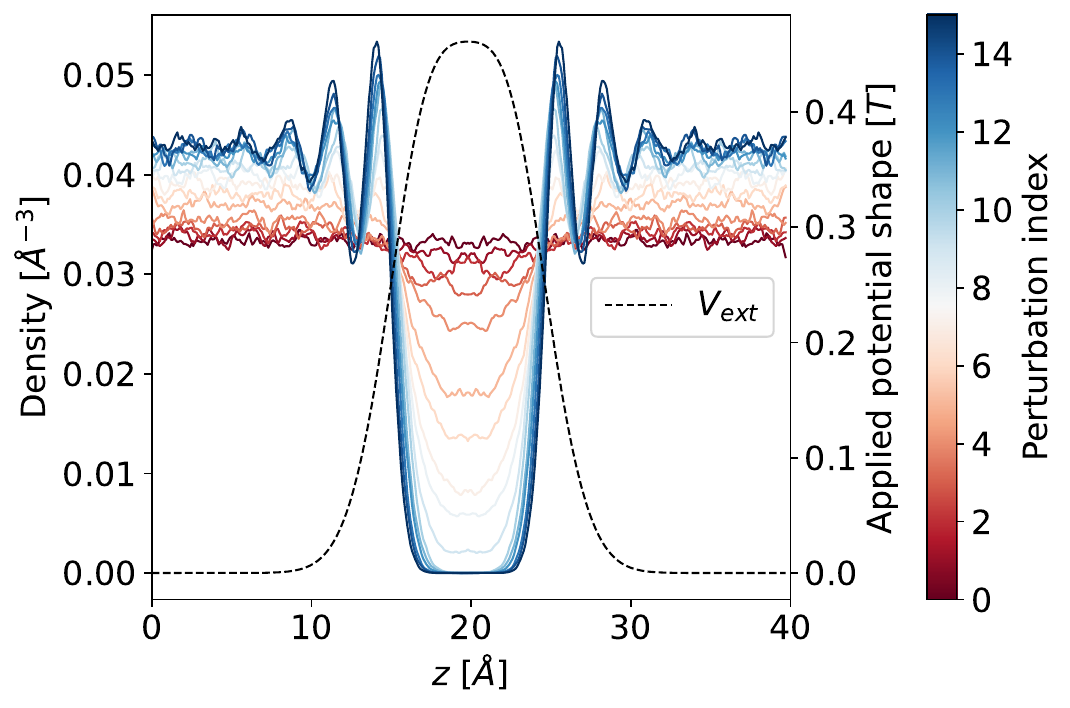}
\caption{Similar to Fig.~\ref{fig:data_hardrod} but for liquid water simulated using SPC/E molecular dynamics simulations.
The simulations are 3D with 1D inhomogeneities, and the density shown is planarly-averaged.
With increasing potential strength, the density profiles of water transition from uniform to strongly inhomogeneous, ranging from complete exclusion of the density to an oscillating shell structure in the high-density region.}
\label{fig:data_spce}
\end{figure}

Figure \ref{fig:data_spce} shows the resulting sequence of (planarly-averaged) density profiles of liquid water from the MD simulations for one external potential shape.
As the strength of the potential increases, the water is excluded from the repulsive region of the potential and develops an oscillating shell structure with increasing magnitude in the attractive regions of the potential.
Note that the density is noisy because the results are derived from histogramming MD trajectories as opposed to evaluating a functional. Nonetheless, an additional challenge for the ML functional is to train an accurate model despite this noise.

\section{Results \& Discussion}

\subsection{Hard-rod fluid}

Because an analytic density functional for the hard-rod fluid exists, this example primarily acts as a proof-of-concept to validate our approach.
After training the ML models until the test loss in energy and potential stop decreasing, we go on to evaluate the models by solving the Euler-Lagrange equation in several external potentials.
This evaluation is a more stringent assessment than the model's test loss, as the variational optimization can find any instabilities in the functional, such as non-positive-definiteness for specific density profiles, and exploit them to make the energy $\to -\infty$.

\begin{figure}
\includegraphics[width=\columnwidth]{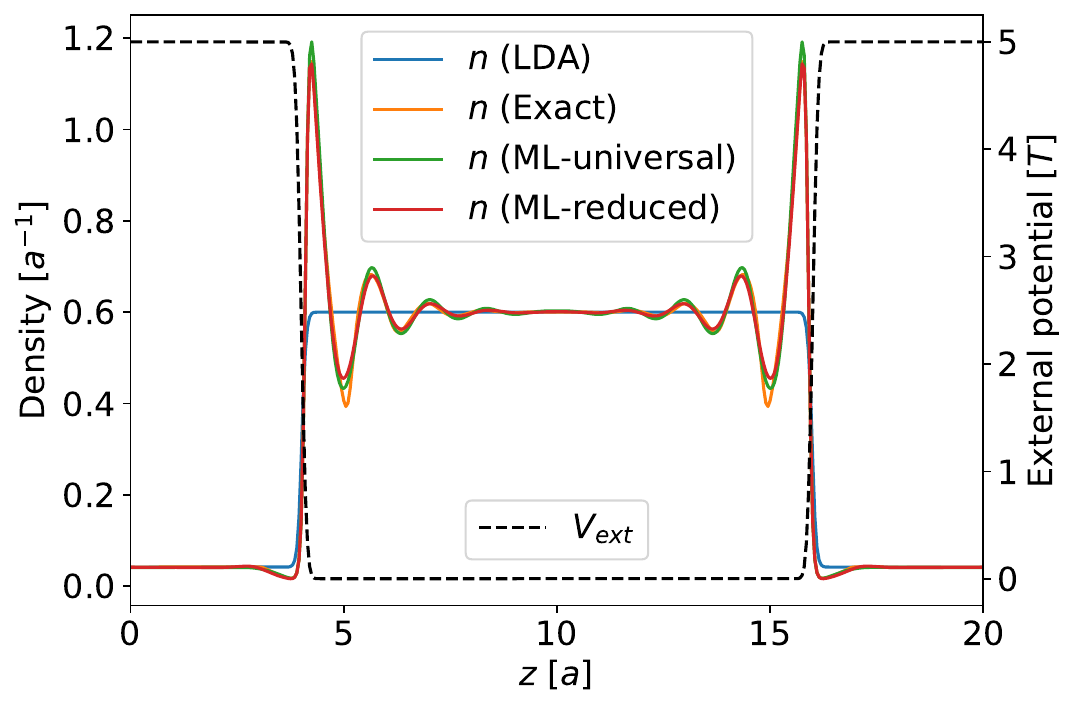}
\caption{Density profiles for the hard-rod fluid confined to an external potential (dashed line) obtained by solving the Euler-Lagrange equation, compared between the exact functional (Eq.~\eqref{eq:hardrods}), its local-density approximation and two ML models.
The LDA is qualitatively incorrect, while both ML models compare excellently to the exact functional, even with highly reduced model size.}
\label{fig:hardrod}
\end{figure}

Figure~\ref{fig:hardrod} shows the density profiles for the hard-rod fluid in a confining potential compared between the local density approximation (LDA), the exact functional, and two ML models, the universal ML model described in Sec.~\ref{sec:loss+training}, and a highly reduced ML model with two convolution layers having just 4 weight functions each (see Table~\ref{tab:hyperparams}). Note that the results for the hard-rod fluid are extremely insensitive to the hyperparameters of the ML model. 
(We discuss hyperparameter selection in more detail with our first non-trivial example in the next section for the electronic exchange energy.)
The density profiles from both ML functionals are in excellent agreement with the exact solution, apart from a smoothing out of the density profiles. The exact functional effectively has a sharp step function as a weight function, which cannot be exactly reproduced by the smooth reciprocal space weight functions of our ML models.
(The physical cases we care about have smooth interactions anyway, unlike the idealized hard-rod model.)
In contrast to the ML model, the LDA result is qualitatively wrong as it misses the shell structure of the fluid entirely.
The performance of the reduced ML model is comparable to our universal ML model, despite having significantly fewer parameters, demonstrating that the even modestly sized ML models suffice for the simple case of the hard-rod fluid.

\subsection{Hartree-Fock exchange energy}

\begin{figure}
\includegraphics[width=\columnwidth]{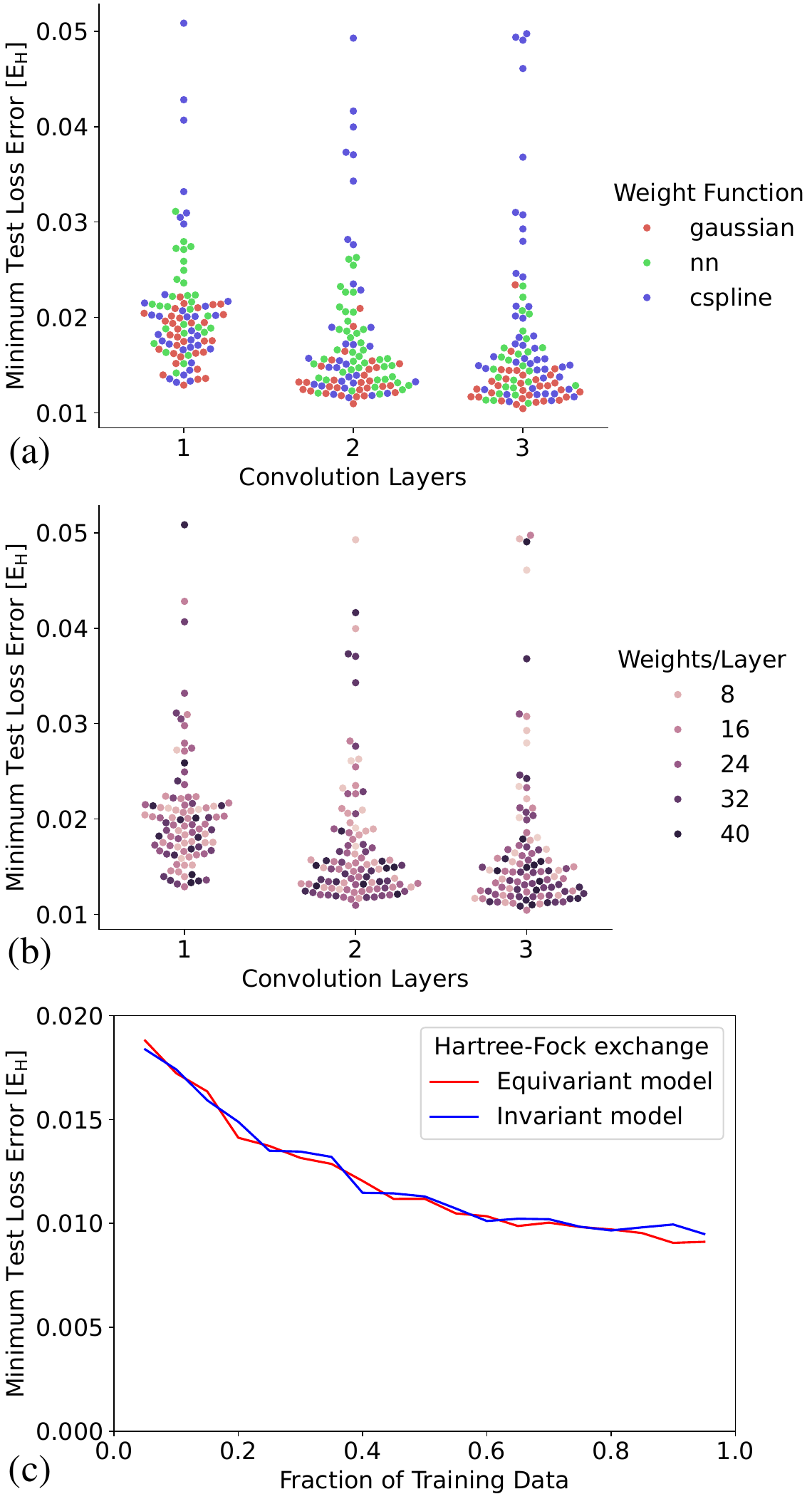}
\caption{Test loss for Hartree-Fock exchange (energy and potential) as a function of the number of convolution layers colored by (a) type of weight function and (b) number of weight functions per layer.
Each point in (a,b) corresponds to a randomly selected set of hyperparameters in the ML model. (c) Test loss for Hartree-Fock exchange as a function of the fraction of data ($\sim9000$ total files) used for training the universal ML model while using the same 80-20 train-test split in all cases. The blue curve shows the test loss of the invariant ML model and the red curve shows the equivariant ML model.}
\label{fig:hypersearch}
\end{figure}

Our first non-trivial example system examines the exact exchange energy of a Hartree-Fock system of electrons in 1D.
We first examined the model performance by varying all hyperparameters of the model, including the number of readout layers, the number of neurons per readout layer, the number of convolution layers, the number of weight functions per convolution layer, weight function type, and the specific hyperparameters within each weight function type (such as $\sigma$ and degree $d$ for the Gaussian weight functions from Eq.~\eqref{eq:gaussweights}). 
We conducted a random search over all combinations of these parameters and trained 300 separate models to the same data (using the same 80-20 train-test split).

Figure~\ref{fig:hypersearch} reveals that the ML model performance is strongly influenced by the number of convolution layers and the type of weight function applied in the ML model. In particular, the accuracy of the model improves a considerable amount increasing from one to two convolution layers and a marginal amount from two to three convolution layers. We also report the test loss as a function of the amount of data used in the training of our universal ML model in Fig.~\ref{fig:hypersearch}(c). The convergence of the test loss error indicates that the amount of data entering the model is sufficient. Figure.~\ref{fig:hypersearch}(c) additionally compares the results from training an equivariant versus invariant ML model, where the invariant model uses only even weight functions. In 1D, we find no discernible difference between the equivariant and invariant ML models, but this result needs to be tested in 2D and 3D before a general conclusion can be made.

\begin{figure}
\centering
\includegraphics[width=0.95\columnwidth]{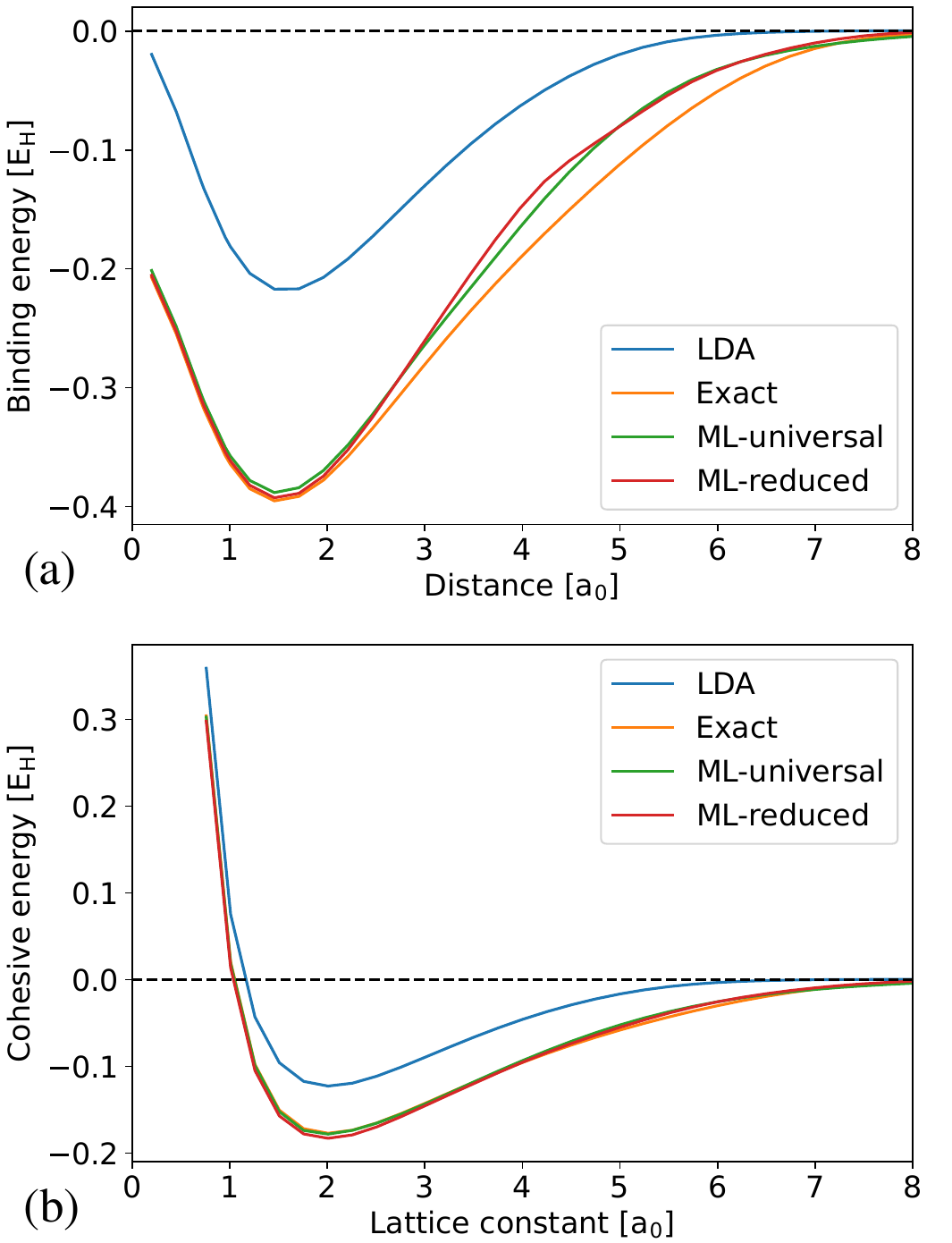}
\caption{(a) Binding energies of a 1D `hydrogen' molecule and  (b) cohesive energies of a infinite hydrogen chain as a function of atom separation, predicted by Hartree-Fock theory compared to the LDA and two ML models of exchange.
The LDA underestimates the binding energies in both cases, while the two ML models, including one with a reduced size, agree remarkably well with the exact results.}
\label{fig:hf-H2}
\end{figure}

Of the three weight function types implemented for this analysis, the Gaussian weight functions perform the best and $\sim 20$ weight functions appear to be sufficient for this system.
For hyperparameters specific to Gaussian weights, the best performing models are obtained with polynomials of degree one (see Eq.~\eqref{eq:gaussweights}).
Other findings from this hyperparameter search show evidence of 2--3 readout layers with $\sim 30$ neurons per readout layer to be sufficient.
We point out the only most noteworthy results of the hyperparameter search here; see the SI for further details and conclusions of this search.
Based on this search, we train a `reduced' model consisting of three convolution layers each with only eight Gaussian weight functions, in addition to the universal hyperparameters from Sec.~\ref{sec:loss+training}. Table~\ref{tab:hyperparams} reports a complete list of all model hyperparameters.

To test the functional for its intended application, energy differences and electronic structure of atomic configurations, we investigate the binding energy of an isolated hydrogen molecule and the cohesive energy of a periodic chain of hydrogen atoms, both computed as the difference in energy from an isolated hydrogen atom.
Note that this 1D atomic `hydrogen' here refers to a species having a nuclear charge of $Z=1$, with a potential based on the regularized 1D Coulomb interaction, and is fundamentally distinct from the actual element hydrogen.
Figure~\ref{fig:hf-H2} shows that the LDA strongly underestimates both the binding and cohesive energies, which is different from the 3D case as a result of the short-range cutoff in the 1D Coulomb kernel that is not present in the 3D kernel. 
Both ML functionals agree remarkably well with the exact solution for both the binding and cohesive energies.
Furthermore the reduced model performs comparably to the universal model, suggesting that the cost of the ML model may be reduced without sacrificing the accuracy of the calculation.

\subsection{Kohn-Sham kinetic energy}

\begin{figure}[t!]
\centering
\includegraphics[width=\columnwidth]{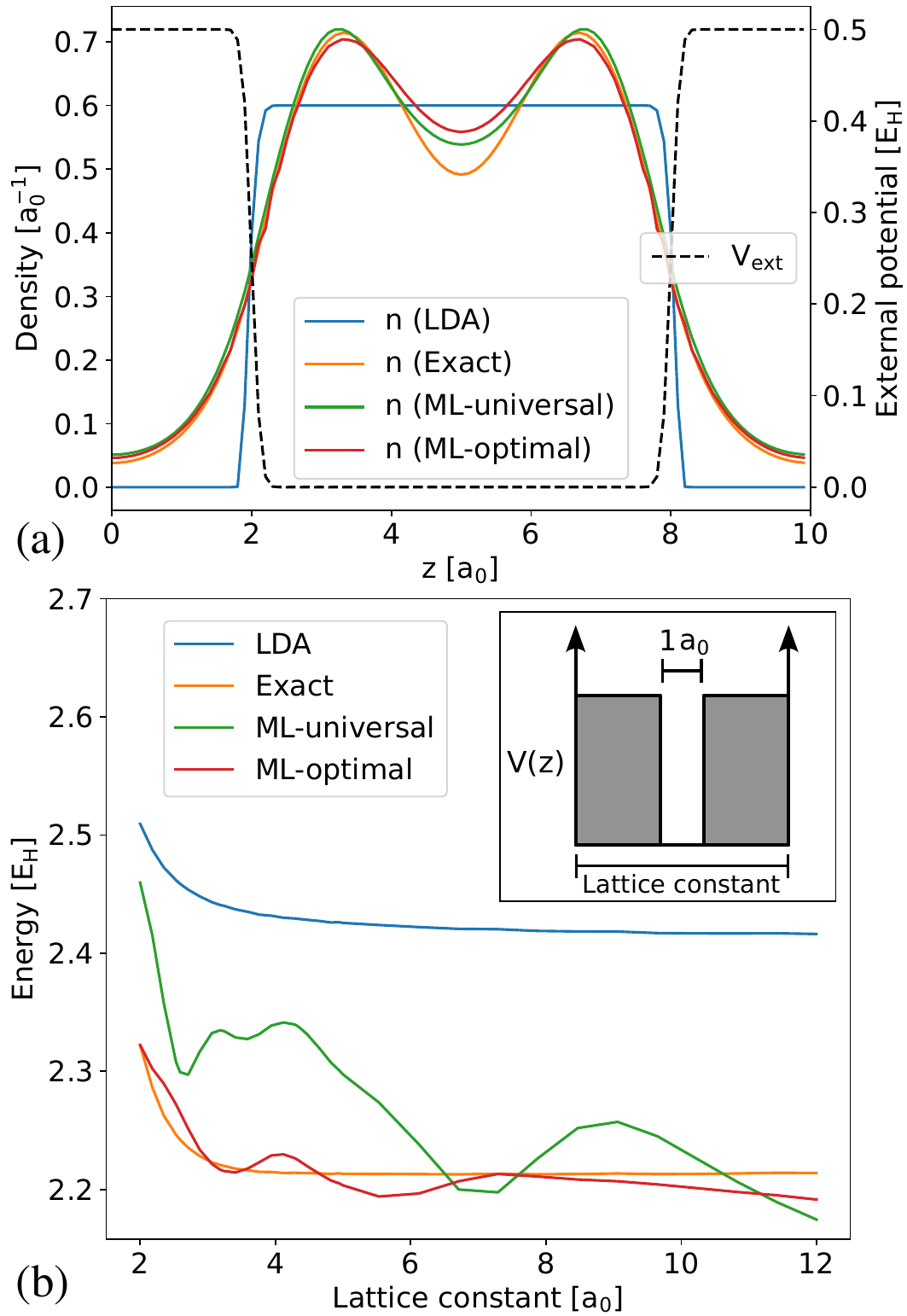}
\caption{(a) Density profiles for non-interacting electrons confined to an external potential (dashed line), predicted by orbital-free kinetic energy functionals at the LDA level and with two ML models, compared to the exact orbital-dependent calculation.
While the LDA result is qualitatively incorrect, the ML models are in reasonable agreement with the exact solution, capturing the tunneling and Friedel oscillations in the high and low potential regions respectively. (b) Total energy predicted by the exact and orbital-free kinetic energy functionals as a function of lattice constant for a periodic well potential illustrated in the inset. Both ML models exhibit qualitatively inaccurate behavior for this test, but the optimal model which has $\sim50$\% more parameters than the universal model produces a more accurate prediction.} 
\label{fig:ks}
\end{figure}

The strongly nonlocal nature of the Kohn-Sham kinetic energy makes this functional especially challenging to train, which we test in another randomized hyperparameter scan of 300 ML models (see the SI).
The hyperparameter search reveals the performance of this functional to be highly sensitive to the choice of hyperparameters, especially compared to the Hartree-Fock exchange energy and the hard-rod fluid.
In particular, we find lower test errors from the larger ML models having more trainable parameters, including both the number of weight functions and convolution layers (see the SI for a more detailed discussion).
Accordingly, we test the optimal (larger) model from this hyperparameter scan along with the universal model.
The optimal model with the lowest test errors consists of two convolution layers with 62 total weight functions each (see Table~\ref{tab:hyperparams} for all other model hyperparameters).

Figure~\ref{fig:ks}(a) shows the resulting density profiles for the Kohn-Sham system in a periodic rectangular well.
The density profiles from both the universal and optimal ML models agree well with the exact solution, while the Thomas-Fermi LDA result yields results that are qualitatively incorrect.
Most impressively, the ML models capture the exponential decay of the electron density in the classically-forbidden tunneling region, as well as the Friedel oscillations in the electron density in the allowed region, all from an orbital-free DFT with no Schr\"odinger equation solution.
Figure~\ref{fig:ks}(b) shows the corresponding energies as a function of spacing between a sequence of periodic wells of the same shape.
Here, the optimal model having a larger number of parameters performs significantly better than the universal model, but the spurious oscillations in the predicted energy as a function of spacing in both ML models indicate that there is further room for improvement.
Additional training data and larger models will therefore likely be necessary for the application of the kinetic energy ML functionals for orbital-free density-functional theory.
Finally, note that these potential wells are qualitatively different from the random periodic potentials in the training set, indicating that our universal ML model is able to generalize well to physically-distinct environments.

\subsection{Liquid water}

Finally, we test the classical DFT for a real fluid, liquid water, with 1D inhomogeneities.
Conducting another hyperparameter scan of 300 random ML models, we determine the optimal set of parameters for water by training to the SPC/E water data described in Sec.~\ref{sec:waterdata}.
We identify a reduced model, having three convolution layers with 18 Gaussian weights each and two readout layers with 80 neurons each, that is moderately smaller than our universal model (see Table~\ref{tab:hyperparams}). 

\begin{figure}
\includegraphics[width=\columnwidth]{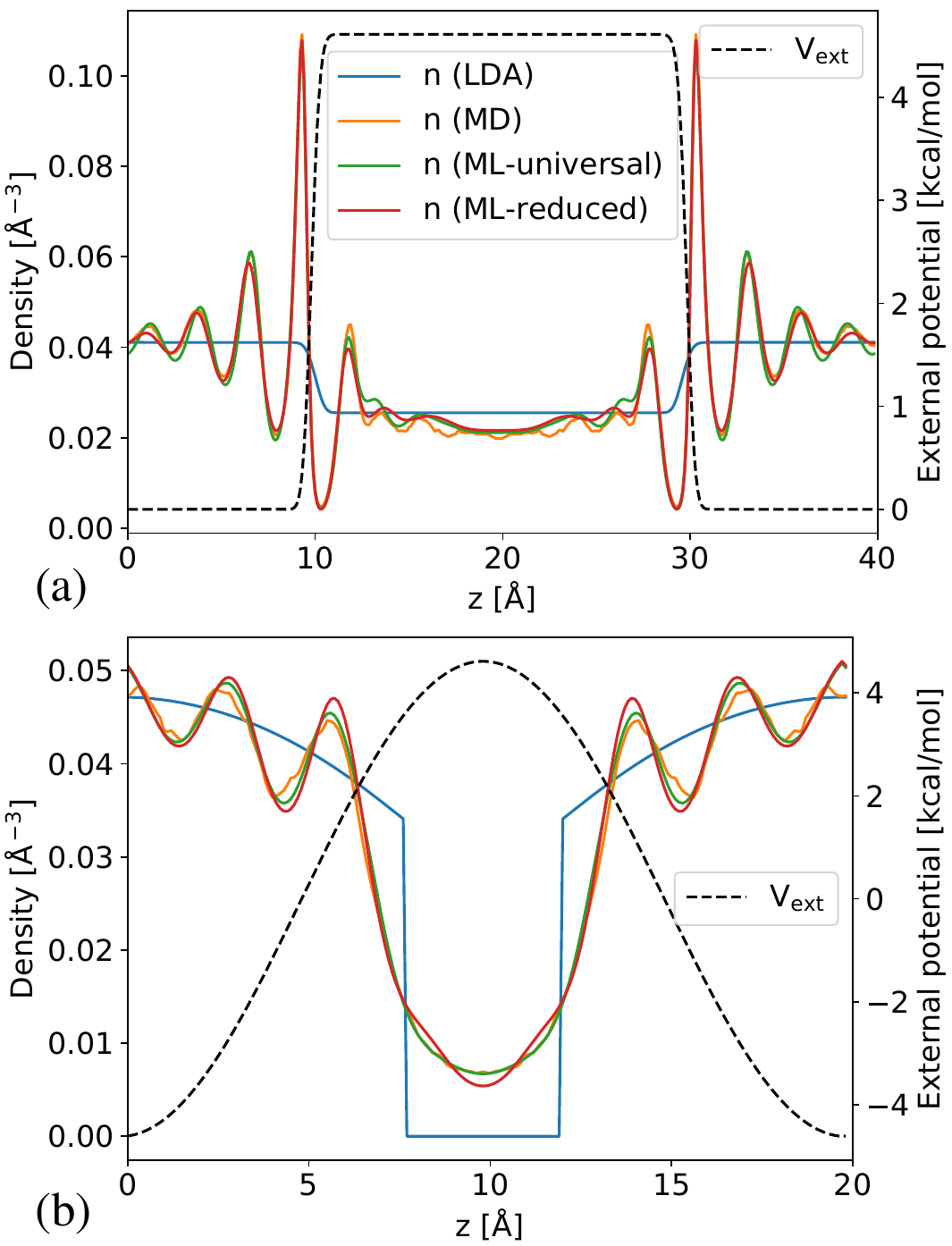}
\caption{Density profiles of liquid water from SPC/E molecular dynamics compared to the LDA and two ML functionals for external potential shapes (dashed lines) completely distinct from those in the training and testing datasets. 
While the LDA is qualitatively inadequate as an excess functional, both ML functionals quantitatively reproduce the non-trivial density profiles with shell structure, in both the (a) square-wave potential and (b) sine-wave potential.}
\label{fig:spce}
\end{figure}

We assess the performances of the reduced and universal ML models as excess functionals for liquid water in classical DFT and solve the Euler-Lagrange equation with two external potential shapes: a square wave and a sine wave, both which are distinct from the potential shapes included within our ML training and testing datasets. 
Figure~\ref{fig:spce} shows that both the optimal and universal ML models successfully reproduce the density profiles from the MD simulations, apart from small quantitative differences which are notably comparable to the magnitude of noise in the MD density.
The density profiles from the LDA, however, are qualitatively incorrect as they miss the oscillations of the density entirely.
These results demonstrate that the universal ML functional is indeed capable of extrapolating well to potentials having abruptly and slowly-varying shapes that are qualitatively distinct from the the training data.
Additionally, the universal ML model predicts the free energy with errors of 0.04 and $<$0.01~kcal/mol/\AA\super{2} respectively for the square and sine wave cases, compared to 0.1 and 0.2~kcal/mol/\AA\super{2} respectively for the LDA, showing that the ML classical DFTs are capable of achieving chemical accuracy for solvation.

\begin{center}
\begin{table*}[t]
\caption{Machine-Learned model hyperparameters}
\begin{tabular}{lcccccc}
\hline
\hline
Model Parameters & Universal model & \begin{tabular}[c]{@{}c@{}}Hard rods\\ (reduced)\end{tabular} & \begin{tabular}[c]{@{}c@{}}Ising model\\ (reduced)\end{tabular} & \begin{tabular}[c]{@{}c@{}}Hartree-Fock\\ (reduced)\end{tabular} & \begin{tabular}[c]{@{}c@{}}Kohn-Sham\\ (optimal)\end{tabular} & \begin{tabular}[c]{@{}c@{}}Liquid Water\\ (reduced)\end{tabular} \\
\hline
Convolution\\
\quad\quad Layers & 3 & 2 & 1 & 3 & 2 & 3 \\
\quad\quad Weights/layer & 20 & 4 & 5 & 8 & 62 & 18 \\
Weights (Gaussian)\\
\quad\quad Gaussian: degree & 1 & 1 & 1 & 1 & 2 & 2 \\
\quad\quad Gaussian: $\sigma_\textrm{max}$ & 2.0--4.0 a.u.\footnote{All models use $\sigma_\textrm{max}=4$ in their respective units with the exception of liquid water which uses $\sigma_\textrm{max}=2.0$~\AA} & 1.0 $a$ & 4 lattice sites & 4.0 $a_0$ & 7.0 $a_0$ & 7.0 \AA \\
Readout\\
\quad\quad Layers & 3 & 3 & 3 & 2 & 3 & 2 \\
\quad\quad Neurons/layer & 100 & 30 & 30 & 90 & 90 & 80 \\
\hline
Trainable parameters & 25301\footnote{The universal model for the Ising and Kohn-Sham cases use an additional 100 parameters to respectively incorporate temperature and local density information in the first readout layer. See discussion around Eq.~\eqref{eq:readout} for more details.} & 2113 & 2116 & 9715 & 39839 & 11105 \\
\hline
\end{tabular}
\label{tab:hyperparams}
\end{table*}
\end{center}

\subsection{Summary}

Once trained, a single universal ML model performs remarkably well across physically distinct problems spanning classical and electronic DFT (demonstrated by the four examples above and a fifth example in the SI).
In comparison to the universal ML model, hyperparameter searches for single-task ML models reveal that significantly smaller ML models are sufficient for all but one case: the Kohn-Sham kinetic energy, which we find to be the most challenging functional to learn. We report the hyperparameters associated with the universal ML model and each of the single-task ML models in Table~\ref{tab:hyperparams}.
The ML functionals effectively reproduce the exact density-profiles in each case, whereas the LDA is qualitative incorrect in all but the exchange energy case, where it is only quantitatively incorrect, as the orbital-dependent kinetic energy regulates the system to produce qualitatively correct behavior.

These results confirm that ML functionals have the potential to substantially improve functional approximations for DFT---for both classical and electronic theories---and expand the scope of DFT to highly non-trivial cases requiring nonlocal functionals.
Hence, ML functionals can expand DFT significantly beyond Kohn-Sham electronic DFT, where only the exchange-correlation functional is approximated and where semi-local approximations have been reasonably successful. Moreover, these same ML functionals can considerably accelerate the usability of classical DFT, where semi-local approximations have been shown to be grossly inadequate, to help overcome the bottleneck of developing classical density functionals.

\section{Conclusions}

We have formulated a general ML functional approximation framework that combines the best features from equivariant convolutional neural networks and the weighted-density approximation. Our approach includes a general protocol for generating data and for training such functionals, which we go on to apply to fundamentally distinct systems.
We systematically tested hundreds of ML models having random combinations of weight function parameterizations, convolution layers, activation functions and readout functions to identify a set of hyperparameters that work for widely-different density functional theories.
We demonstrated that we can build a universal ML model, that is trained to data from these distinct physical systems---but uses exactly the same hyperparameters---that performs remarkably well and reproduces highly non-trivial features in the density profiles and energies that are missed completely by semi-local approximations.
In particular, we showed that in addition to exactly solvable models, the ML functional performs excellently for the exact exchange energy and kinetic energy of electrons in 1D, as well as for the excess functional of (3D) liquid water with 1D inhomogeneity.
Remarkably, these functionals show promise of achieving useful orbital-free electronic DFT as well as classical DFT of liquid water with chemical accuracy for solvation.

Here, we prototyped all models in 1D \edited{with open-source software implementations} in order to rapidly generate data for several physically-distinct systems and test the universality of the ML approach.
While we {theoretically formulate} the extension to \edited{2D and} 3D for the ML functional approach already, \edited{this will require extension of the software implementations to higher dimensions.
Additionally,} the cost of generating high quality data to train the functionals can become the bottleneck, \textit{e.g.}, OEP calculations for exchange in 3D are very expensive, and collecting 2D and 3D density profiles from MD will require much longer collection times due to increased statistical errors; these extensions may require further methodological developments beyond brute-force large-scale computation. 
\edited{Similarly, developing unified functionals for a combined classical and electronic treatment of solvated systems using joint density-functional theory\cite{JDFT} is an ambitious task -- in any dimension -- and will hinge on the ability of reliably generating data for such systems.}
Finally, for the classical DFT case, this framework can eliminate the need for classical interatomic potentials altogether by applying ML interatomic potentials trained to \emph{ab initio} molecular dynamics (AIMD) of inhomogeneous fluids,\cite{mdext} directly constraining classical DFT to AIMD for truly \emph{ab initio} solvation models.

\section{Supplementary Material}
The supplementary material includes the link to open-source git repository containing the software used to produce and test machine-learned density functionals, additional options for readout layer choice and weight function types, results for the 1D inhomogeneous Ising model, and results from our hyperparameter search analysis.

\section*{Acknowledgements}
This work was supported by the U.S. Department of Energy, Office of Science, Basic Energy Sciences, under Award No. DE-SC0022247.
Calculations were carried out at the Center for Computational Innovations at Rensselaer Polytechnic Institute, and at the National Energy Research Scientific Computing Center (NERSC), a U.S. Department of Energy Office of Science User Facility located at Lawrence Berkeley National Laboratory, operated under Contract No. DE-AC02-05CH11231 using NERSC award ERCAP0020105.

\section*{References}
\bibliography{references}
\end{document}